%% file: main.tex
\documentclass[10pt, conference]{IEEEtran}
\let\labelindent\relax

\input{packages}
\input{macros}

\makeatletter
\def\subsubsection{\@startsection{subsubsection}
                                 {3}
                                 {\z@}
                                 {0ex plus 0.1ex minus 0.1ex}
                                 {0ex}
                                 {\bfseries\normalsize}}
\makeatother

\begin{document}

\restoresymbol{change}{comment}

\title{Shipwright: A Human-in-the-Loop System \\ for Dockerfile Repair}

\newcommand{\UWMad}[1][ersity]{Univ#1 of Wisconsin--Madison} 

\author{\IEEEauthorblockN{Jordan Henkel\IEEEauthorrefmark{1}\textsuperscript{\textsection},
Denini Silva\IEEEauthorrefmark{2}\textsuperscript{\textsection},
Leopoldo Teixeira\IEEEauthorrefmark{2}, 
Marcelo d'Amorim\IEEEauthorrefmark{2}, and
Thomas Reps\IEEEauthorrefmark{1}}
\IEEEauthorblockA{\IEEEauthorrefmark{1}\UWMad{}, Madison, WI, USA}
\IEEEauthorblockA{\IEEEauthorrefmark{2}Federal University of Pernambuco, Recife, PE, Brazil\\
\{jjhenkel,reps\}@cs.wisc.edu\quad\{dgs,lmt,damorim\}@cin.ufpe.br}}

\maketitle
\begingroup\renewcommand\thefootnote{\textsection}
\footnotetext{Equal contributions.}
\endgroup

\thispagestyle{plain}
\pagestyle{plain}

\begin{abstract}

Docker is a tool for lightweight OS-level virtualization.\Ignore{through images and containers.} Docker images are created by performing a build, controlled by a source-level artifact called a \dfile{}.
We studied \dfile{}s on GitHub, and---to our great surprise---found that over a quarter of the examined \dfile{}s failed to build (and thus to produce images).
To address this problem, we propose \tname, a human-in-the-loop system for finding repairs to broken \dfile{}s.
\tname\ uses a modified version of the BERT language model to embed build logs and to cluster broken \dfile{}s.
Using these clusters and a search-based procedure, we were able to design \numrepairs{} rules for making automated repairs to \dfile{}s. 
With the aid of \tname{}, we submitted \totalNumberOfPRs{} pull requests (with a \totalPectenOfPRsAccepted{}\% acceptance rate) to \github\ projects with broken \dfile{}s. Furthermore, in a ``time-travel'' analysis of broken \dfile{}s that were later fixed, we found that \tname{} proposed repairs that were equivalent to human-authored patches in \numMatchesPerc{}\% of the cases we studied.
Finally, we compared our work with recent, state-of-the-art, static \dfile{} analyses, and found that, while static tools detected possible build-failure-inducing issues in \binnacleMaybeCoversP{}--\hadolintMaybeCoversP{}\% of the files we examined, \tname{} was able to detect possible issues in \coverageOfAllFilesIncludingSuggestions{}\% of the files and, additionally, provide automated repairs for \coverageOfAllFiles{}\% of the files.
\end{abstract}



\section{Introduction}
\label{sec:intro}
Docker is one the most widely used tools for virtualization. With $\sim$79\% of IT companies using it~\cite{portworx}\Ignore{ and over 3.7 million unique installations of the VS Code Docker extension~\cite{vscode:Docker}\Mar{reader may wonder why VS Code installation is relevant}}, Docker has made an impact on developers' day-to-day work. Developers use Docker to author images via an artifact called a \emph{Dockerfile}. Images can be based on a variety of operating systems, but primarily, Docker images are Linux-based. \dfile{}s are, effectively, a linear sequence of ``setup instructions'' that tell the Docker engine how to prepare an image. The final built image is then used to run Docker containers. These containers are similar to lightweight virtual machines. Each container starts with the clean environment specified by its originating Docker image. Together, images and containers allow for isolation, scaling, and reproducibility.

Nonetheless, we \added[id=J]{found} that \added[id=T]{over 26\%} of the analyzed samples of \dfile{}s obtained ``in the wild'' (sourced from GitHub) \added[id=J]{failed} to build successfully. 
This \added[id=T]{high a percentage} is surprising, because it runs counter to one of the core tenets of Docker, namely, reproducibility.
Furthermore, it is outside of the scope of recent efforts \added[id=T]{to} \added[id=L]{statically} \added[id=T]{analyze Dockerfiles} to detect such failures. For example, Hadolint \cite{github:Hadolint} can detect mistakes such as missing or incorrect flags---\myeg{}, forgetting the use of \CodeIn{--assume-yes/-y} when invoking \CodeIn{apt-get install} in a \dfile{}. This flag is required because a \dfile{} build runs without interaction; therefore, forgetting this flag may cause the build to hang. 
Unfortunately, while Hadolint can detect such a mistake statically, many breakages occur due to a \emph{change in the external environment} and \emph{not a change in the source \dfile{}}. \added[id=J]{These observations add to the mounting evidence that external changes, such as dependencies changing, can often lead to broken build-related artifacts \cite{broken-snapshots-jsep, HireBuild}.}
\added[id=J]{As further evidence of this trend,} our prototype tool, \tname{}, was used to guide \totalNumberOfPRsAccepted{} accepted pull requests on \github~and, for each of these patches, the underlying issue was caused by a \emph{change in \added[id=T]{the} external environment} \added[id=J]{(see \Cref{sec:brk-changes-ext-files} for an example of one such change)}.

\begin{standout}
{\textbf{The Problem:}} Over a quarter of the GitHub repositories with \dfile{}s that we analyzed had a broken \dfile{}. Current state-of-the-art (static) analysis for \dfile{}s is largely incapable of detecting and/or \added[id=J]{repairing} such broken \dfile{}s. 
\end{standout}

\added[id=J]{Given this situation, our work seeks to meet the following high-level goal:}

\begin{standout}
{\textbf{Our Goal.}} \added[id=J]{Aid developers in automatically} repairing broken \dfile{}s, with the hope of reducing the high percentage of broken \dfile{}s that we observed on GitHub.
\end{standout}

\subsection{Contributions} 



Our work makes the following contributions: 1)~\emph{Technique}.~We introduce a human-in-the-loop approach to fixing broken \dfile{}s;~2)~\emph{Tool}.~We made available a tool implementing our technique; and 3)~\emph{Data}.~We made available an extension of the \texttt{binnacle} dataset \cite{henkel2020dataset}, including build logs.


\vspace{0.1cm}\noindent\Contrib{Technique.}~Unlike previous approaches that attempt to mine patterns automatically and directly from \dfile{}s, \tname{} follows a human-in-the-loop approach to build a \added[id=L]{\emph{repair} database}. We include human supervision to broaden the effectiveness of \tname{}: a fully automatic approach would limit the scope of repairs that could be detected. \tname{} is designed to act as a ``co-pilot'' that can side-step the limitations of a completely automatic approach \added[id=J]{with the goal of constructing a} comprehensive database of \added[id=L]{repairs}. \added[id=L]{To build such a database,} \added[id=J]{\tname{} uses  clustering, human supervision, and a search-based recommendation system we built to leverage
\deleted[id=T]{the} vast community-knowledge bases, such as StackOverflow and Docker's community forum. In general, we require repairs to incorporate both some kind of pre-condition (pattern) and a transformation function (patch).} 

\added[id=J]{During clustering, one challenge that \tname{} needs to address is the heterogeneity of the data}, which is a mixture of code and natural language. \added[id=J]{The mixing of code and natural language makes it non-trivial to design a featurization method for clustering. Therefore, to address this challenge, \tname{} uses a modified version of Google's BERT model \cite{reimers-2019-sentence-bert, BERT_2019} to \emph{embed Dockerfile build logs}. By using a pre-trained transformer-based neural model, we are able to side-step tedious feature engineering, and benefit from the diverse corpora \added[id=T]{on which BERT-based models have been trained.}} Using this embedding, we perform clustering with \texttt{HDBSCAN} \cite{hdbscan_original}, in the vector space of embedded build logs, and use the results to cluster failing builds.
By using the vectors generated through BERT, we leverage the ``understanding'' encoded in BERT's language model.
To our surprise, we found that recent off-the-shelf language models work well in this domain. \added[id=J]{Using the generated clusters, we employ human supervision to intuit, for each cluster, a likely root cause of failure and, if possible, engineer one or more automated repairs to save to \tname{}'s database. Later, \tname{} will use this human-generated repair database to attempt automatic repair of failing \dfile{}s.}

\vspace{0.1cm}\noindent\Contrib{Tool.}~The scripts to automate each of the steps of \tname{} (see Figure~\ref{fig:workflow}) are publicly available at \toolUrl{}.

\vspace{0.1cm}\noindent\Contrib{Data.}~We have identified a subset of \dfile{}s from the \texttt{binnacle} dataset \cite{henkel2020dataset} that are amenable to automated builds.
(The dataset and filtering criteria are described in~\Cref{sec:dataset}.)
We have built these files \emph{in-context} (an expensive operation that requires hundreds of hours of compute time), and captured detailed data from the results, including logs from the builds. These build logs represent a significant expansion of the data in the original \texttt{binnacle} dataset, and it is our hope that this extended data will accelerate research on diagnostic tools for \dfile{} analysis and repair. 
\added[id=J]{This expanded data is available at \artifactUrl.}

\subsection{Evaluation}

We evaluated several aspects of \tname{};
a summary of our results is as follows:
(i) Broken \dfile{}s are prevalent: in the data we analyzed, \breakageRate{} of \dfile{}s failed to build.
(ii) Even using optimistic criteria, existing static tools are capable of identifying
the cause of a failure in \added[id=T]{only} \binnacleMaybeCoversP{}\%--\hadolintMaybeCoversP{}\% of \added[id=J]{the broken \dfile{}s.}
(iii) \tname{} is capable of clustering broken \dfile{}s and offering actionable solutions: for files that clustered, \tname{} provides \added[id=J]{automated} repairs in \numPercRepair{}\% of the cases; for files that did not cluster, \tname{} is still able to provide \added[id=J]{automated repairs} in \numRepairNotInClustersPerc{}\% of cases.
(iv) \added[id=J]{In a ``time-travel'' analysis,} we found that \tname\ would be able to provide actionable solutions to \percCoverageRepairConfirmation\% of the \dfile{}s that we found initially broken and then \added[id=J]{subsequently} fixed in their respective repositories.
\added[id=J]{Finally}, we used the reports from \tname\ to submit \totalNumberOfPRs{} Pull Requests to still-broken \dfile{}s.
Of these, \totalNumberOfPRsAccepted{} have been accepted.
These results provide initial, yet strong, evidence that \tname\ is useful to help developers fix broken \dfile{}s. 

\newcommand{\rqone}{How prevalent are \dfile{} build failures in projects that use \docker\ on \github? Can existing (static) tools identify the failure-inducing issues within these broken files?}
\newcommand{\rqtwo}{\added[id=J]{Can we use off-the-shelf language models, like BERT, to easily cluster broken \dfile{}s?}}
\newcommand{\rqthree}{\added[id=J]{How effective is \tname{} in producing repairs? (i) To what extent do repairs cover the failures from our dataset? (ii)~For failures that can be clustered, is it possible to generalize repairs? (iii)~What can be done for failures from non-clustered files?}}
\newcommand{\rqfour}{How effective is \tname{} in reducing the number of broken \dfile{}s in public repositories?}




\section{Sources of Build Failures}
\label{sec:example}

This section describes \added[id=J]{some of the} distinct sources of problems that can lead to build failures in \dfile{}s.

\subsection{Breaking Changes in External Files}
\label{sec:brk-changes-ext-files}

\added[id=J]{This kind of failure occurs when a \dfile{}'s external dependencies (such as the file's base image or URLs embedded within the file) are changed; often these changes external to the \dfile{} will require a change in the \dfile\ itself.} To illustrate this problem, consider the case where the developer used \CodeIn{latest} to indicate the version of the base image of her \dfile{}, as in \CodeIn{FROM ubuntu:latest}. 
These base images are downloaded from Docker Hub~\cite{dockerhub}, a distributed database that is part of the \docker{} ecosystem. 
The problem with using \CodeIn{latest} is that a change to the base image may require changes to the \dfile{}. Unfortunately, there is no clear way to incorporate those required changes automatically. 
For instance, the \CodeIn{python-pip} package is part of Python 2, and Python 2 is unavailable on Ubuntu images higher than 18.04.
Consequently, a build on a \dfile{} with the command \CodeIn{apt-get -y install python-pip} will pass when the file is based on Ubuntu images 18.04 and lower, but it will fail on higher versions, including the latest LTS version of Ubuntu. 

We used the \tname{} toolset to analyze \added[id=J]{and cluster} hundreds of broken \dfile{}s, looking for common error patterns in their build logs and associated \dfile{}s. \added[id=J]{Using a human-in-the-loop process, we then extracted\deleted[id=M]{ possible} patterns and associated them with candidate repairs.}
For example, when running the command \CodeIn{docker build} on the \dfile{} \CodeIn{FROM ubuntu:latest...RUN apt-get -y install python-pip...}, \docker{}\replaced[id=M]{ reports the message}{ produces the string} ``\textit{Unable to locate package python-pip}'' on output. When that\deleted[id=M]{output } message is present in the logs, we found that the typical image in \dfile{}s is Ubuntu and the version is either undefined, \added[id=L]{\CodeIn{latest}}, or \added[id=L]{\CodeIn{20.04}}. We also noticed that this error message appears not only with \CodeIn{python-pip}, but with other packages \added[id=L]{as well}.

We expressed these patterns with regular expressions to be checked against the \dfile\ (the static data) and error logs (the dynamic data). For instance, we expressed the pattern for the problem above with the regex ``\CodeIn{FROM ubuntu}($\epsilon$ | :latest | :20.04)'' $\wedge$ ``\textit{Unable to locate package} (.*)'' $\in$ \underline{log}. Note that such \added[id=T]{an} expression 1)~defines a pattern over the \dfile, 2)~defines a pattern over the output log, and 3)~uses the groups ``($\epsilon$...)'' and ``(.*)'' to bind data to variables for later use.

\begin{figure}[t!]
\begin{lstlisting}[style=Bash]
# solution 1, use version 18.04
FROM ubuntu:18.04
RUN apt-get -y install python-pip
... #remaining code

# solution 2, manually install the package
FROM ubuntu:latest
ARG DEBIAN_FRONTEND=noninteractive
RUN apt-get -y install python2 curl software-prop... \
  && add-apt-repository universe \
  && curl https://.../get-pip.py --output get-pip.py \
  && python2 get-pip.py
... #remaining code
\end{lstlisting}
    \vspace{-2ex}
    \caption{\label{fig:ex1}Solving \CodeIn{python-pip} unavailable on \CodeIn{ubuntu:latest}.}
    \vspace{-3ex}
\end{figure}

\deleted[id=J]{We used that pattern to search on the web for solutions to build problems.\Leo{Technically, we did not use the regular expression pattern to search the web, is that a problem?}}
\Cref{fig:ex1} shows two possible solutions to the problem above.
The first solution is to fix the base image to the most recent version \replaced[id=T]{with which}{where} the command can still be executed. The following abstract operation characterizes the repair: \underline{replace} \CodeIn{\$0} \underline{with} \CodeIn{:18.04}. The symbol \CodeIn{\$0} refers to the regex group matching the Ubuntu image in the \dfile\ (\ie{}, ``($\epsilon$...)'' ) that must be replaced with one specific Ubuntu version (\myeg{}, \CodeIn{:18.04}). 
The second solution is to install Python 2 and its toolset. The following operation  characterizes the repair: \underline{add} \CodeIn{ARG DEBIAN\_FRONTEND=\dots} \underline{after} ``\CodeIn{FROM ubuntu}($\epsilon$ | \added[id=L]{\CodeIn{:latest}} | \added[id=L]{\CodeIn{:20.04}})''. 
The symbol ``\dots'' is \added[id=T]{a} placeholder for the text associated with the second solution from \Cref{fig:ex1}.

\tname{} records the association between a given pattern (of build error) and possible solutions, such as the two repairs above. From this information, \tname\ is able to repair broken \dfile{}s whose build logs match some of the pre-recorded patterns. \Cref{table:fixes-main-2} describes this example under the row with ``Id'' 5. Note that the repair operation consists of multiple possible solutions, hence the comma and dots, as we only show the first solution. In this case, \tname{} produces two versions of the \dfile{} and the developer should choose which one suits best their needs. We elaborate on \tname's workflow in the following sections.

\deleted[id=J]{This problem happened in \NumCasesOfIdFiveInDataset{} of the \numFilesInClusters{} broken \dfile{}s...} 

It is worth noting that prior work has investigated the impact of breaking changes in package-management systems (\myeg{}, in the Linux package manager, npm, maven, \etc)~\cite{6606587,mccamant-ecoop2004,breaking-changes-node,4222580,4019575}. \tname\ is not restricted to this kind of issue\added[id=T]{,} and \deleted[id=T]{it} is distinct from prior work on the application context and solution used. Section~\ref{sec:related} elaborates on related work.

\subsection{Inconsistent Version Dependency Within Project}

This kind of failure occurs when there is an inconsistency between the versions of a \dfile\ and some of the files it \added[id=J]{references} within the project. 
Figure~\ref{fig:ex-broke-dockerfile} shows a concrete example that illustrates the problem. The \dfile{} requires an image for Ruby version 2.6.3, whereas the application code declares a dependency on a newer Ruby version~(2.6.5)~\cite{dependency-rubylate}. The execution of the command \CodeIn{RUN bundle install} triggers an error, producing the following message on output ``Your Ruby version is 2.6.3, but your Gemfile specified 2.6.5''. In this case, the solution was to replace line 1 of this \dfile{} with \CodeIn{FROM ruby:2.6.5}. The pattern and corresponding repair explained above are listed in Table~\ref{table:fixes-main-2} under row ``Id'' 6. A similar repair is ``Id'' 1, which is also related to Ruby.

\begin{figure}[t!]
\begin{lstlisting}[style=Bash]
FROM ruby:2.6.3
RUN apt-get update -qq && apt-get install -y \
...
RUN gem install bundler:2.0.1
RUN bundle install # <-- Gemfile depends on ruby 2.6.5! 
ADD . /app 
\end{lstlisting}
    \vspace{-2ex}
    \caption{Inconsistent Ruby version dependencies.}
    \label{fig:ex-broke-dockerfile}
    \vspace{-3ex}
\end{figure}

\subsection{Missing Commands in the \dfile}

This failure occurs when a given \dfile{} uses a command that is unavailable on a given image. The solution in that case is to install the command using the proper syntax, \replaced[id=T]{because}{as} it depends on the version of the image. Table~\ref{table:fixes-main-2} lists one example of this error pattern and the respective repair under row ``Id'' 8. 

\subsection{Project-Specific Failures and Suggestions}
We observed that many broken \dfile{}s require repairs that are project-specific and \replaced[id=T]{cannot}{can't} be generalized. In those cases, \tname\ is unable to produce a repair to the broken file. Instead, in those cases, \tname\ provides \emph{suggestions}. For instance, consider the case where a \dfile{} includes a command to deploy a Node.js server, such as \CodeIn{RUN npm run build}. The execution of that command fails because there is an error in the Node.js project. There is nothing to fix within the \dfile{}. The developer needs to analyze what is wrong in her Node.js project and fix it.
In that case, \tname\ reports a suggestion\added[id=T]{,} such as ``NPM build error..''.
As another example, consider \replaced[id=T]{a case in which}{the cases where} a command refers to a broken link, such as \CodeIn{RUN wget <\emph{url}>}. \tname\ \replaced[id=T]{cannot}{can't} guess how to fix the broken link. Table~\ref{table:fixes-main-1} shows a sample of suggestions provided by \tname{}. \added[id=J]{In the future, it would be interesting to examine how one might combine \tname{} with other tool\added[id=T]{-specific} and language\added[id=T]{-}specific repair techniques.
Tools that have a \emph{combined understanding} of DevOps artifacts and the programs these artifacts support represent an intriguing area for future work.}

\begin{figure*}[t]
    \centering
    \includegraphics[width=\textwidth]{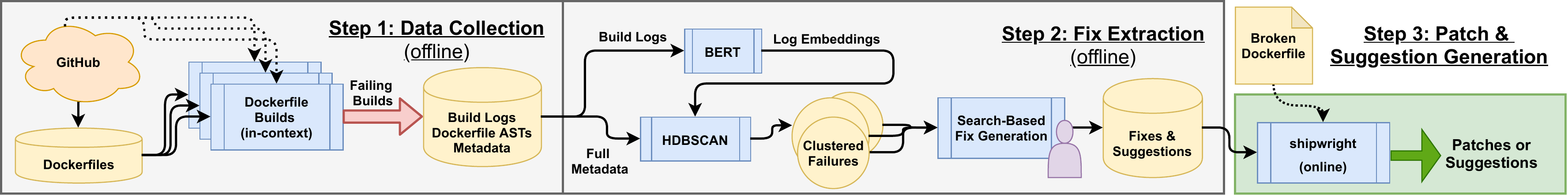}
    \caption{\tname{}'s 3-step workflow. In step (1), a database of \dfile{}s and GitHub metadata is used to perform \emph{in-context} builds. The results of these builds are stored in a local database along with various forms of metadata. In step (2), \tname{} uses BERT and HDBSCAN to cluster build data~\cite{reimers-2019-sentence-bert, BERT_2019, hdbscan_original}. The clusters are then fed to \tname{}'s search-based repair-generation and suggestion-generation process. During this step, \tname{}, with the assistance of a human, builds a database of repairs and suggestions. Finally, in step (3), online usage begins: new files are fed to \tname{} and, if matching database entries are found, \tname{} provides relevant repairs or suggestions.
    }
    \label{fig:workflow}
    \vspace{-3ex}
\end{figure*}

\section{Dataset}
\label{sec:dataset}

We use an expanded version of the \binn\ dataset~\cite{henkel2020dataset} as the source of \dfile{}s to analyze. \added[id=J]{The \binn\ dataset consists of \emph{all} \dfile{}s from GitHub repositories \emph{with ten or more stars}. These Dockerfiles represent a broad and largerly unfiltered picture of the state of Dockerfiles one might find in popular GitHub repositories.} Although the original dataset was created in 2019, the \binn\ toolchain allows us to capture recent data using the same methodology\added[id=J]{; thus, we populated our dataset with more recent data (June 2020) extracted using the same tools.} Unfortunately, directly using the \dfile{}s in this dataset is challenging for two reasons: (i) many \dfile{}s in the dataset come from the same repository and, in such cases, the \emph{purpose} of the \dfile{}s is obscured, making efforts---like automated builds---more difficult; (ii) many \dfile{}s are nested deep within repositories (especially when repositories contain many independent projects or services). In either case, automated builds are challenging because the \emph{intent} behind the \dfile{} is difficult to infer. In case~(i), it is difficult to infer which of the (many) \dfile{}s should be built.
In case~(ii), it is difficult to infer an appropriate \emph{context directory}, which is a pre-requisite to building a Docker image.

\vspace{0.1cm}\noindent\textbf{Dataset Filtering.} To address these issues, we filtered the files from the \binn\ dataset using two criteria: (a) we only considered repositories with \emph{a single \dfile{}}, and (b) we required that the given \dfile{} \replaced[id=L]{resides}{reside} within the repository's \emph{root directory}.
For such a repository, it is not unreasonable to assume two things: (i) the \dfile{} is intended to produce an artifact corresponding to the given repository (because it is the \emph{only} \dfile{} in that repository), and (ii) the \dfile{} likely uses the repository's root directory as its \emph{build context}: the \dfile{} resides in the root directory, and the \CodeIn{docker build} command assumes, by default, that the target \dfile{} resides within the given context directory. \added[id=J]{Performing this filtering yielded \numTotalFiles{} repositories and corresponding \dfile{}s that may be amenable to automated builds. It is this subset of the original dataset (a refreshed version of \binn{}'s dataset) that we \replaced[id=T]{used in our studies}{will explore in further detail}.}

\vspace{0.1cm}\noindent\textbf{Building Dockerfiles at Scale.} \added[id=J]{\tname{} performed \emph{in-context} builds on \numFilesBuilt{} of the \numTotalFiles{} \dfile{}s in our \emph{filtered} dataset (recall: we filter to find Dockerfiles from the original dataset that are likely amenable to \emph{automated} builds). Although we would have liked to use all \numTotalFiles{} \dfile{}s, we encountered some problems performing in-context builds on that many files in a reasonable time frame, which forced us to use a smaller set of \numFilesBuilt{} files. 
We tried various approaches \replaced[id=T]{to}{of} scaling these in-context builds, but distributing this kind of process would require Docker installations across a wide variety of machines---in practice, this requirement was difficult to satisfy: none of the distributed-computing resources we had access \replaced[id=L]{to}{too} would allow us to control a \emph{Docker daemon} on the distributed machines (while, for many distributed platforms, running \emph{containers} is easy). This requirement, while understandable from a security perspective, forced us to run builds in a non-distributed way (on a single large server). Given this constraint, some builds either \emph{time out} (we set a limit of 30 minutes) or, due to contention from running multiple builds on this single server and daemon instance, some builds fail to complete due to errors internal to the Docker daemon; builds that fail in this manner are marked as \emph{undetermined}. Instead of revisiting undetermined builds, we spent our resources on building a larger portion of our filtered dataset.
Through this process, we capture\added[id=T]{d the} results from \replaced[id=T]{\numFilesBuilt{} \dfile{}}{tens of thousands of \dfile{}} builds\added[id=T]{,} and save\added[id=T]{d} these results to a database for further analysis.
\Cref{sec:data-collection} describes this offline data processing with the \tname{} toolset in further detail.}

\begin{figure*}[t]
    \centering
    \includegraphics[width=\textwidth]{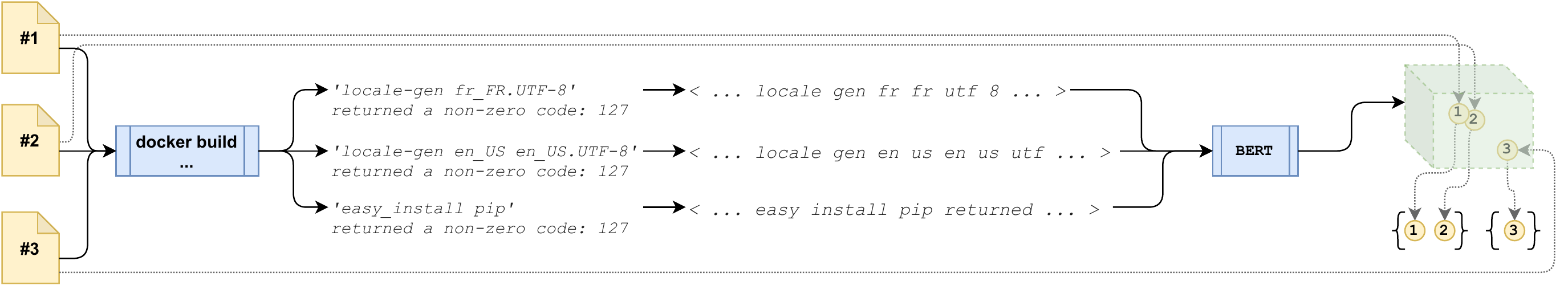}
    \caption{\textbf{Clustering Example.}
    Starting with several broken Dockerfiles, \tname{} clusters by extracting standard error logs, applying aggressive token splitting \added[id=J]{(we split on snake and camel case, as well as several operators that may occur in code snippets)} and string normalization \added[id=J]{(we lowercase the input, clear large blocks of repetitive characters, like extraneous white space, and we strip certain special characters/unicode)}, and passing the resulting sequences to BERT. BERT takes the input sequences and produces vectors (shown as points in a high-dimensional space). \tname{} uses that mapping, from failing build logs to points in a vector space, along with a clustering algorithm (HDBSCAN), to produce its clusters, shown in the bottom right of the figure. \added[id=J]{The final clustering process has the advantage of being \emph{semantics aware}: BERT understands some of the nuances of the English language, and our clustering benefits from this capability.}}
    \label{fig:clustering-ex}
    \vspace{-3ex}
\end{figure*}

\section{\tname}
\label{sec:approach}

\Cref{fig:workflow} shows the workflow of \tname\ as a pipeline of three steps, organized according to their respective goals.

The goal of the first step is to analyze a corpus of broken \dfile{}s---mined from \github{}---and to perform \emph{in-context} builds so that logs can be acquired (\Cref{sec:data-collection}). The goal of the second step is to cluster broken \dfile{}s and find repairs (\ie{}, transformation functions on \dfile{}s) \deleted[id=J]{or suggestions} (\Cref{sec:clustering}). 
Given a cluster, \tname\ automatically elaborates search queries from log files of representative \dfile{}s within the cluster. 
A human then supervises the creation of repairs and suggestions by (i)~looking for error patterns as manifested in existing QA forums resulting from the search query, and 
(ii)~creating plausible repairs \added[id=J]{(or, if no automated repairs are possible, creating suggestions)} which are saved in a database for later, online, use. 
\deleted[id=J]{Suggestions cover the cases where repairs require domain knowledge (e.g., fixing a compilation error).}
Finally, the third (interactive) stage of the pipeline looks for actual repairs for a broken \dfile~(\Cref{sec:repairing-files}). This component takes as input the output produced in the previous (offline) stages and a broken \dfile, and \added[id=J]{produces either (i) an automated repair, (ii) a suggestion (in cases where repair cannot be automated), or (iii) an indication that no existing repairs or suggestions apply.} The following sections describe each step in detail.

\subsection{Data Collection}
\label{sec:data-collection}
This component of \tname{} builds and analyzes \dfile{}s from a pre-existing corpus. We utilize the \numTotalFiles{} \dfile{}s described in \Cref{sec:dataset} as our input corpus\replaced[id=M]{. Those}{ because those} files were filtered to be amenable to automated builds. For each file in this corpus, \tname{} does the following: 1) \emph{Clones} the originating repository for the given \dfile{} into a unique \CodeIn{/tmp/<repo-id>} directory; 2) \emph{Runs} \CodeIn{docker build -f <Dockerfile> /tmp/<repo-id>}, which builds a \CodeIn{<Dockerfile>} from our dataset using the root directory of the cloned repository as the build context. Building \emph{in context} is crucial because the build may need to access files from the originating repository to complete successfully\replaced[id=M]{. Although}{, and although} we are interested in build failures, we want to avoid trivial failures; 3) \emph{Discards} builds that still have trivial failures; and 4) \emph{Saves}, for each failing build, execution logs (standard out\added[id=M]{put} and standard error streams), the AST for the given \dfile{}, and various metadata (\myeg{}, repository information, image history, and a log of the git clone procedure). \replaced[id=T]{An example would be a broken build caused by an execution failure in a directive, such as}{For example, broken builds that were caused by execution failures in directives like} \CodeIn{COPY} or \CodeIn{ADD}. Although builds are performed \emph{in context}, it is still possible that the Dockerfile is intended to be built as part of a more complex workflow with a context directory that is different from the repository root directory. Unfortunately, this information may exist in some third-party script, or may be user-supplied.


\Cref{fig:workflow} illustrates this workflow in the \added[id=M]{box named }``Step 1: Data Collection.''\deleted[id=M]{box} We note that data collection is quite costly: we ran a 32-core CentOS workstation for several weeks and, during that time, managed to build about two thirds (\numFilesBuilt{}) of the \numTotalFiles{} files in our dataset. \added[id=J]{(See \Cref{sec:dataset} for a discussion of the difficulties of building many \dfile{}s at scale.)} Although builds can be parallelized, there is only one Docker daemon per installation of Docker---this situation creates a limit to the practical concurrency that can be achieved, along with the network bandwidth to the workstation used for analysis. We attempted to perform builds on a high-throughput cluster but, unfortunately, the strict security requirements of such clusters prevent deploying a workflow involving Docker images builds, which effectively execute untrusted code.


\subsection{Repair \& Suggestion Extraction}
\label{sec:clustering}

This step of \tname{} works as follows. First, it uses \added[id=J]{HDBSCAN}, applied to embeddings,\footnote{Embeddings refer to high-dimensional vectors of numbers that are used as a proxy for non-numeric artifacts (such as text or code). Embeddings are often of use, because many operations can be performed in the resulting vector space, and later mapped back to the originating artifacts.} to partition
the \dfile{} data produced in the previous step. Second, it uses those clusters to assist a human with the task of searching for solutions and building a database of repairs\deleted[id=J]{ and suggestions}. We now elaborate on each of these steps.

\subsubsection*{Clustering}
\tname{} attempts to cluster failing \dfile{} builds using embeddings and \texttt{HDBSCAN} (a hierarchical variant of  \texttt{DBSCAN}, a classic clustering algorithm \cite{dbscan_original}).
The difficulty of clustering in this domain is two-fold:
(i) the data to cluster is heterogeneous, and is often a mix of code and natural language (\ie{}, the build logs, which will often contain a description of the failure in English and a reproduction of the Bash or \dfile{} snippet that leads to the error), and
(ii) although we would like to cluster on the \emph{cause} of build failures, we do not have a way to definitively extract the cause of a given build failure;
therefore, we must use data that is, at best, a proxy or symptomatic of the root cause of failure.
In particular, we use a tokenized version of the build logs for the failing build \added[id=J]{(which may include a variety of things: debug output, warnings, and errors---some of which may include \replaced[id=L]{code snippets}{snippets of code})} as input to BERT to produce embeddings.

Despite these challenges, \tname{} is able to perform clustering by leveraging a key insight:
recent off-the-shelf language models, such as BERT, GPT-2 and, recently, GPT-3, have reached impressive levels of sophistication \cite{BERT_2019,radford2019language,brown2020language}.
Given the inputs these models are trained on (roughly, massive crawls of the internet), it is highly likely that such models have seen websites like StackOverflow, which mix both natural language and code.
Therefore, to address challenge (i) (the heterogeneous mix of code and natural language), \tname{} leverages a sufficiently sophisticated off-the-shelf language model (BERT),\footnote{Ideally, one would try GPT-3 because it is the newest and largest such language model; unfortunately, for now, access to GPT-3 is quite limited, and we were unable to obtain access.} to obtain \emph{embeddings}.
In particular, \tname{} uses a variant of BERT suited to the task of sentence embedding, in which similar sentences should end up ``close'' in the embedding space.
\tname{} applies this BERT variant to the last few lines of the \added[id=J]{error} logs to produce a vector representation of each broken \dfile{}.
These vectors are then fed to HDBSCAN, which produces clusters.
\Cref{fig:clustering-ex} illustrates the \tname{} clustering process.

\subsubsection*{Searching for Repairs or Suggestions}
\label{sec:fixes}
This component of \tname\ takes a set of clusters as input, and produces a \emph{list of pairs} as output. The first element of the pair is a signature that identifies the issue (in the \dfile\ and its logs) whereas the second element \added[id=M]{of the pair} is either (i) a repair, consisting of a pure transformation function that takes a \dfile\ as input and produces another file as output, or (ii) a suggestion (about what needs to be repaired and how) for the cases where human knowledge is necessary to prepare the repair.  We elaborate on each of these two cases in the following\added[id=J]{ and \Cref{table:fixes-main-1,table:fixes-main-2} provide examples of such pairs.}

\textbf{Case 1 (Searching for Repairs):} \tname\ uses a search-based recommendation system
to assist a human in locating repairs of broken \dfile{}s. \tname\ proceeds as follows:
it selects a cluster and a representative \dfile{} from that cluster;
it extracts keywords from the logs of that file;
it builds a search string from those keywords;
it submits the corresponding query to a search engine;
it filters the outputs from related community forums;
and it reports a list of the top-5 URLs as output for a human to inspect.
Human inspection consists of reading proposed solutions on discussion forums, and then applying a given solution to the representative \dfile\ from the cluster. If that solution is plausible, \ie{}, if it allows the \dfile\ to build an image successfully, the next step is to check if the error-pattern/repair-function pair is applicable to other \dfile{}s in the cluster. While doing so, the human inspector \replaced[id=M]{looks}{should look} for opportunities to generalize the pattern and repair\deleted[id=M]{ the} function to avoid overfitting a solution to a particular case. For instance, in the example given in \Cref{sec:brk-changes-ext-files}, the initial solution was too narrow, focusing on fixes of files containing the exact message ``\CodeIn{Unable to locate package python-pip}'' in the output log. However, we observed similar error messages, referring to different packages. In this case, the solution was to replace ``\CodeIn{python-pip}'' in that string with a symbolic name for a package. To sum up, \tname\ leverages community knowledge bases (\myeg{}, StackOverflow and Docker's community forums) to find solutions to known issues, such as those presented in \Cref{sec:example}.

\input{figs/fix-repairs-table}

\tname{} supports a total of \numrepairs{} repair patterns.
Table~\ref{table:fixes-main-2} shows the pairs of (1) build-error signatures---referred to as a pattern---and (2) a corresponding repair for 10 of them. Column ``Id'' shows the id of a pair. Column ``Pattern'' shows the error pattern, which is a regular expression that matches a string in the error logs (\added[id=M]{the }dynamic part) and/or a string in the \dfile{} (\added[id=M]{the }static part).
Column ``Repair'' shows a function, in natural language, describing how to transform and fix a broken \dfile{}.
We use the keywords \underline{add}, \underline{remove}, and \underline{replace} to describe operations that need to be performed on the \dfile{}.
We informally described the semantics of these operations with examples in \Cref{sec:brk-changes-ext-files}.
Although there is no fundamental reason preventing us from creating these transformations automatically, we wrote the code implementing these transformation functions because we found empirically that creating these functions \replaced[id=L]{was}{were} not a time-consuming error-prone task.
Finally, column ``Src.'' shows\replaced[id=M]{ a reference for the solution}{ a solution documented}  on the web.

\input{figs/fix-recommendations-table}

\textbf{Case 2 (Searching for Suggestions):} There are cases where \tname\ cannot produce a repair.
For example, a \dfile\ whose build fails because of a compilation error or broken URL requires a human to fix the underlying error.
For those cases, we report a suggestion, \ie{}, generic advice on what needs to be done. 
Table~\ref{table:fixes-main-1} shows a small sample from the total of \numrecommendations{} suggestion patterns that \tname{} supports (in addition to \numrepairs{} repair patterns).
Column ``Id'' shows the id of the suggestions, column ``Pattern'' shows the signature, and column ``Suggestions'' shows the suggestion message.

\subsection{Repair and Suggestion Generation}
\label{sec:repairing-files}

\tname{} can be used to repair \dfile{}s or provide suggestions using the database generated in step~2~(\Cref{sec:fixes}).
Given a \dfile{}, \tname{} iteratively examines the repairs and suggestions and, given a match, it either (i) produces a patched file, by applying a repair, or (ii) provides a suggestion message to the user. If neither a repair nor a suggestion with a matching pre-condition exists within the database, \tname{} is still able to use its search-based process to guide a human in producing fixes and suggestions, as we did in step 2~(\Cref{sec:fixes}). This search-based process provides a user with a small set of (filtered) links to resources likely to help in fixing the given input file. In summary, \tname{}, during step 3, produces either: (i) a \dfile{} repair, (ii) a suggestion on how to fix the broken build, or (iii) a curated set of results from a search-based process that may provide solutions to the underlying build issue. 


\section{Evaluation}
\label{sec:evaluation}

The goal of \tname{} is to help developers fix broken \dfile{}s. It does that through a combination of (i) clustering of broken \dfile{}s (by likely root cause), and (ii) a search-based method to find repairs \added[id=J]{(and, if no automated repairs are feasible, suggestions)}. To gain insights into the landscape of broken \dfile{}s used in \github{} projects and to understand \tname{}'s efficacy, we pose the following research questions.

\subsection*{\textbf{RQ1.} \rqone{}} 

\subsubsection*{Rationale}
The purpose of this question is to evaluate the potential impact of \tname{}. 
If build failures are rare, then impact is limited. 
Furthermore, reproducibility is a core tenet of \docker{}---it would be surprising to find many broken \dfile{}s. 
We also assess the ability of existing (static) tools to identify issues that may  lead to failing \dfile{} builds.


\subsubsection*{Metrics}
We used the following metrics to answer RQ1:~1)~the fraction of \dfile{}s \added[id=J]{in our dataset} with builds that fail; 2)~the relationship between failures and project popularity; and 3)~the success rate of existing (static) tools in predicting \dfile{} build failures. 
The first metric evaluates the fraction of \dfile{}s that we mined from \github\ that fail to build because the \dfile\ is broken (for non-trivial and non-toolchain-related reasons). The second metric examines the relationship between the number of GitHub stars a given repository has (a common proxy for popularity on GitHub) and whether that repository contains a broken \dfile{}. This measurement helps us ascertain whether popular repositories suffer from broken \dfile{}s at the same rate as less popular repositories. Recall that we do not have any \dfile{}s from repositories with less than 10 GitHub stars (\Cref{sec:dataset}). Finally, the third metric ascertains the ability of pre-existing tools, namely, Hadolint \cite{github:Hadolint} and \texttt{binnacle}'s rule checker \cite{henkel2020learning}, to find issues within broken \dfile{}s.

\subsection*{\textbf{RQ2.} \rqtwo{}}

\subsubsection*{Rationale}
Given the number of observed failures, it is reasonable to ask whether many failures are \emph{unique}. \deleted[id=J]{If that were the case, it would be challenging to automatically repair them because the ability to generalize would be limited---there would be many very-specific cases to handle.} If many failures are similar, one might hope that generalized repairs exist. Furthermore, if failing \dfile{} builds can be clustered, those clusters may be used to bootstrap finding \deleted[id=J]{such} repairs. \added[id=J]{Finally, if we can leverage the level of understanding available in large off-the-shelf language models (like BERT), then we can design robust clustering routines with little specialized engineering effort, and avoid techniques based on manually designed heuristics.}

\subsubsection*{Metrics}
\added[id=J]{To answer RQ2, we examine the percentage of clusters (generated using \tname{}), where all elements share a single (likely) root cause.}
This metric provides insight into \tname{}'s ability to cluster broken \dfile{}s and the usefulness of those clusters---good clustering allows for finding multiple exemplars for a single failure which, in turn, makes the task of \added[id=J]{generating automated repairs simpler.}

\subsection*{\textbf{RQ3.} \rqthree{}}

\subsubsection*{Rationale}
The purpose of this question is to evaluate \tname{}'s effectiveness on our dataset. If proposed solutions are unable to cover a variety of \dfile{}s, then \tname{}'s usefulness is questionable.

\subsubsection*{Metrics} 
1)~We measured the fraction of broken \dfile{}s (from our dataset) for which \tname\ \added[id=J]{produces a repair}. 2)~For the \emph{set of clusters} that \tname{} produces, we measured the extent to \added[id=J]{which repairs generalize}. For that, we measure ``coverage'' (i) in the cluster that originated that pattern, and (ii) across different clusters. Coverage refers to the portion of elements within a cluster that match the same pre-condition \added[id=J]{for a repair}.
3)~For broken \dfile{}s that \emph{did not cluster}, we measured how often \tname{} provides \added[id=J]{a repair}. \added[id=J]{In all cases, we also evaluated \tname{}'s ability to provide \emph{suggestions} if repairs were not feasible.}  Collectively, these metrics measure how effective \tname\ is in proposing solutions to the broken files in our dataset.


\subsection*{\textbf{RQ4.} \rqfour{}}

\subsubsection*{Rationale}
Although RQ3 seeks to evaluate \tname{}'s ability to fix broken \dfile{}s, there still remains a question of \tname{}'s usefulness in practice. RQ4 seeks an understanding of \tname{}'s effectiveness to meet our overarching goal: fixing broken \dfile{}s in public repositories.  

\subsubsection*{Metrics}
To answer RQ4, we used two metrics: 1)~What proportion of \dfile{}s that appear in our dataset as broken \dfile{}s, but have since been fixed, \emph{would also have been fixed, had we applied \tname{}}? 2)~How often can we use \tname{} to produce pull-requests that are accepted by external reviewers? 
The first metric refers to a kind of ``time-travel'' analysis because, using updates that took place during the period in which we built \tname{}, we can attempt to measure how successful we \emph{would have been} had \tname{} existed at an earlier date, and had we applied it. Nevertheless, this metric is still a ``simulated'' one. Therefore, the second metric quantifies \tname{}'s ``real-world'' applicability by actually using it to produce repairs and submitting them for (external) review.

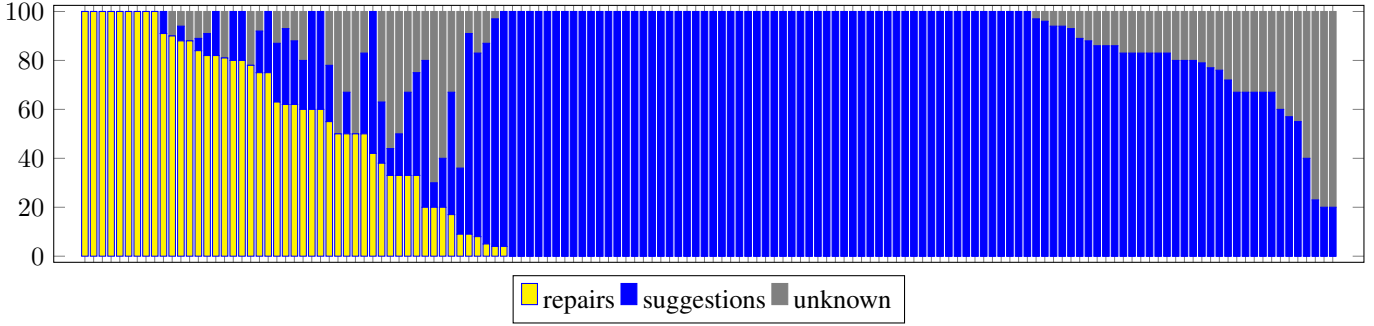
\begin{figure*}
    \centering
    \begin{tikzpicture}
        \begin{axis}[
            ybar stacked,
        	bar width=2.5pt,
            enlargelimits=0.025,
            legend style={at={(0.5,-0.05)}, anchor=north,legend columns=-1},
            symbolic x coords={1, 2, 3, 4, 5, 6, 7, 8, 9, 10, 11, 12, 13, 14, 15, 16, 17, 18, 19, 20, 21, 22, 23, 24, 25, 26, 27, 28, 29, 30, 31, 32, 33, 34, 35, 36, 37, 38, 39, 40, 41, 42, 43, 44, 45, 46, 47, 48, 49, 50, 51, 52, 53, 54, 55, 56, 57, 58, 59, 60, 61, 62, 63, 64, 65, 66, 67, 68, 69, 70, 71, 72, 73, 74, 75, 76, 77, 78, 79, 80, 81, 82, 83, 84, 85, 86, 87, 88, 89, 90, 91, 92, 93, 94, 95, 96, 97, 98, 99, 100, 101, 102, 103, 104, 105, 106, 107, 108, 109, 110, 111, 112, 113, 114, 115, 116, 117, 118, 119, 120, 121, 122, 123, 124, 125, 126, 127, 128, 129, 130, 131, 132, 133, 134, 135, 136, 137, 138, 139, 140, 141, 142, 143, 144},
            xtick=data,
            width=19cm,
            height=5cm,
            xticklabels={,,}
        ]

        \addplot+[ybar, fill=yellow] plot coordinates {(1, 100) (2, 100) (3, 100) (4, 100) (5, 100) (6, 100) (7, 100) (8, 100) (9, 100) (10, 91) (11, 90) (12, 88) (13, 88) (14, 84) (15, 82) (16, 82) (17, 81) (18, 80) (19, 80) (20, 78) (21, 75) (22, 75) (23, 63) (24, 62) (25, 62) (26, 60) (27, 60) (28, 60) (29, 55) (30, 50) (31, 50) (32, 50) (33, 50) (34, 42) (35, 38) (36, 33) (37, 33) (38, 33) (39, 33) (40, 20) (41, 20) (42, 20) (43, 17) (44, 9) (45, 9) (46, 8) (47, 5) (48, 4) (49, 4) (50, 0) (51, 0) (52, 0) (53, 0) (54, 0) (55, 0) (56, 0) (57, 0) (58, 0) (59, 0) (60, 0) (61, 0) (62, 0) (63, 0) (64, 0) (65, 0) (66, 0) (67, 0) (68, 0) (69, 0) (70, 0) (71, 0) (72, 0) (73, 0) (74, 0) (75, 0) (76, 0) (77, 0) (78, 0) (79, 0) (80, 0) (81, 0) (82, 0) (83, 0) (84, 0) (85, 0) (86, 0) (87, 0) (88, 0) (89, 0) (90, 0) (91, 0) (92, 0) (93, 0) (94, 0) (95, 0) (96, 0) (97, 0) (98, 0) (99, 0) (100, 0) (101, 0) (102, 0) (103, 0) (104, 0) (105, 0) (106, 0) (107, 0) (108, 0) (109, 0) (110, 0) (111, 0) (112, 0) (113, 0) (114, 0) (115, 0) (116, 0) (117, 0) (118, 0) (119, 0) (120, 0) (121, 0) (122, 0) (123, 0) (124, 0) (125, 0) (126, 0) (127, 0) (128, 0) (129, 0) (130, 0) (131, 0) (132, 0) (133, 0) (134, 0) (135, 0) (136, 0) (137, 0) (138, 0) (139, 0) (140, 0) (141, 0) (142, 0) (143, 0) (144, 0) };

        \addplot+[ybar,blue] plot coordinates {(1, 0) (2, 0) (3, 0) (4, 0) (5, 0) (6, 0) (7, 0) (8, 0) (9, 0) (10, 9) (11, 0) (12, 6) (13, 0) (14, 5) (15, 9) (16, 18) (17, 0) (18, 20) (19, 20) (20, 0) (21, 17) (22, 25) (23, 24) (24, 31) (25, 26) (26, 20) (27, 40) (28, 40) (29, 23) (30, 0) (31, 17) (32, 0) (33, 33) (34, 58) (35, 25) (36, 11) (37, 17) (38, 34) (39, 42) (40, 60) (41, 10) (42, 20) (43, 50) (44, 27) (45, 82) (46, 75) (47, 82) (48, 93) (49, 96) (50, 100) (51, 100) (52, 100) (53, 100) (54, 100) (55, 100) (56, 100) (57, 100) (58, 100) (59, 100) (60, 100) (61, 100) (62, 100) (63, 100) (64, 100) (65, 100) (66, 100) (67, 100) (68, 100) (69, 100) (70, 100) (71, 100) (72, 100) (73, 100) (74, 100) (75, 100) (76, 100) (77, 100) (78, 100) (79, 100) (80, 100) (81, 100) (82, 100) (83, 100) (84, 100) (85, 100) (86, 100) (87, 100) (88, 100) (89, 100) (90, 100) (91, 100) (92, 100) (93, 100) (94, 100) (95, 100) (96, 100) (97, 100) (98, 100) (99, 100) (100, 100) (101, 100) (102, 100) (103, 100) (104, 100) (105, 100) (106, 100) (107, 100) (108, 100) (109, 100) (110, 97) (111, 96) (112, 94) (113, 94) (114, 93) (115, 89) (116, 88) (117, 86) (118, 86) (119, 86) (120, 83) (121, 83) (122, 83) (123, 83) (124, 83) (125, 83) (126, 80) (127, 80) (128, 80) (129, 79) (130, 77) (131, 76) (132, 72) (133, 67) (134, 67) (135, 67) (136, 67) (137, 67) (138, 60) (139, 57) (140, 55) (141, 40) (142, 23) (143, 20) (144, 20) };

        \addplot+[ybar, gray] plot coordinates 
        {(1, 0) (2, 0) (3, 0) (4, 0) (5, 0) (6, 0) (7, 0) (8, 0) (9, 0) (10, 0) (11, 10) (12, 6) (13, 12) (14, 11) (15, 9) (16, 0) (17, 19) (18, 0) (19, 0) (20, 22) (21, 8) (22, 0) (23, 13) (24, 7) (25, 12) (26, 20) (27, 0) (28, 0) (29, 22) (30, 50) (31, 33) (32, 50) (33, 17) (34, 0) (35, 37) (36, 56) (37, 50) (38, 33) (39, 25) (40, 20) (41, 70) (42, 60) (43, 33) (44, 64) (45, 9) (46, 17) (47, 13) (48, 3) (49, 0) (50, 0) (51, 0) (52, 0) (53, 0) (54, 0) (55, 0) (56, 0) (57, 0) (58, 0) (59, 0) (60, 0) (61, 0) (62, 0) (63, 0) (64, 0) (65, 0) (66, 0) (67, 0) (68, 0) (69, 0) (70, 0) (71, 0) (72, 0) (73, 0) (74, 0) (75, 0) (76, 0) (77, 0) (78, 0) (79, 0) (80, 0) (81, 0) (82, 0) (83, 0) (84, 0) (85, 0) (86, 0) (87, 0) (88, 0) (89, 0) (90, 0) (91, 0) (92, 0) (93, 0) (94, 0) (95, 0) (96, 0) (97, 0) (98, 0) (99, 0) (100, 0) (101, 0) (102, 0) (103, 0) (104, 0) (105, 0) (106, 0) (107, 0) (108, 0) (109, 0) (110, 3) (111, 4) (112, 6) (113, 6) (114, 7) (115, 11) (116, 12) (117, 14) (118, 14) (119, 14) (120, 17) (121, 17) (122, 17) (123, 17) (124, 17) (125, 17) (126, 20) (127, 20) (128, 20) (129, 21) (130, 23) (131, 24) (132, 28) (133, 33) (134, 33) (135, 33) (136, 33) (137, 33) (138, 40) (139, 43) (140, 45) (141, 60) (142, 77) (143, 80) (144, 80) };

        \legend{\strut repairs, \strut suggestions, \strut unknown, }
        \end{axis}
    \end{tikzpicture}
    \caption{\label{fig:proportion}Proportion of different kinds of solutions within each cluster (excluding singleton clusters).
    }
    \vspace{-3ex}
\end{figure*}

\subsection{Answering RQ1: \rqone}

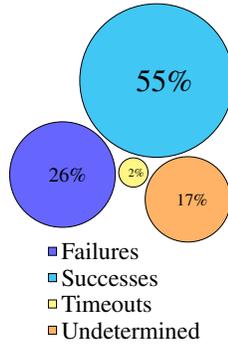
\begin{wrapfigure}{R}{0.35\linewidth}
\vspace{-6ex}
\resizebox{\linewidth}{!}{
    \begin{tikzpicture}
    \pie[cloud,text=legend,scale font]{26/\huge{Failures}, 55/\huge{Successes}, 2/\huge{Timeouts}, 17/\huge{Undetermined}}
    \end{tikzpicture}
}
\setlength{\abovecaptionskip}{-5pt}
\setlength{\belowcaptionskip}{-10pt}
\caption{Breakdown of the \numFilesBuilt{} files we attempted to build.}
\label{fig:breakdown-pie}
\end{wrapfigure}

To answer RQ1, we used \tname{} to build a random sample of \dfile{}s from our (filtered) dataset. In total, we tried to build \numFilesBuilt{} \dfile{}s and found \numFilesBroken{} broken \dfile{}s. This gives us an estimated \breakageRate{} ``breakage rate'' for \dfile{}s in our overall dataset. The large amount of broken \dfile{}s on GitHub runs counter to one of the core reasons for using Docker: \emph{reproducibility}. Aside from broken \dfile{}s, we encountered \numFilesTimedout{} \dfile{}s with builds that time out (we use a threshold of 30 minutes) and \numFilesUndet{} \dfile{}s with undetermined results (which arise due to the pressure that multiple concurrent builds place on the Docker daemon). Neither timeouts nor builds with undetermined results are counted as broken \dfile{}s. Instead, we count these results as successful builds to give a conservative estimate (and lower bound) of the ``breakage rate'' for \dfile{}s in our dataset. \Cref{fig:breakdown-pie} provides a visual overview of these categories.

To put these results in context, we also examined the distribution of stars for the \numFilesBroken{} repositories in our dataset.
For these repositories, we find that: (i) a third have 18 stars or fewer, (ii) a third have greater than 18 stars, but fewer than 51 stars, and (iii) a third have 51 or more stars. This distribution was surprising, especially because some repositories with broken \dfile{}s had many thousands of stars. We spot-checked some of these cases and found that, indeed, even quite popular repositories can have broken \dfile{}s. For example, the MEAN stack project~\cite{mean-stack} has over 12K stars, yet it contains a \dfile{} that fails to build. 

Finally, we also tested the capabilities of two existing (static) tools: \texttt{binnacle} \cite{henkel2020learning} and Hadolint \cite{github:Hadolint}. For both tools, we sought an estimate of the number of broken \dfile{}s for which each tool \emph{identifies a possible build-breaking issue}.
Because we found it impractical to \replaced[id=L]{manually examine}{examine manually} the tools' outputs on each of the \numFilesBroken{} broken files, we instead used a (generous) estimate based on how often each tool reports a rule violation for an issue that \emph{might cause a build to break}.
For example, Hadolint can identify when the version of an image used for a base in a \dfile{} is un-pinned;
thus, if Hadolint reports a rule violation in this category, on any file, we count it as Hadolint identifying a \emph{possible build-breaking issue} (and mark the file as ``solved'' by Hadolint).
In total, Hadolint identifies such issues in only \hadolintMaybeCoversP{}\% of files, and \texttt{binnacle} identifies such issues in only \binnacleMaybeCoversP{}\% of files.




\begin{standout}
    \textbf{Summary of RQ1:} The presence of broken \dfile{}s on \github{} is common. Furthermore, even highly starred repositories sometimes contain broken \dfile{}s. Finally, existing static tools only identify plausible build-breaking issues in \binnacleMaybeCoversP{}--\hadolintMaybeCoversP{}\% of cases (and, even when issues are identified, such tools do not provide repairs). 
\end{standout}

\subsection{Answering RQ2: \rqtwo}

\added[id=J]{To generate the clusters we use throughout our evaluation, we first performed a grid search against \tname{}'s clustering algorithm. During this search, we focused on exploring the space of hyperparameters used in HDBSCAN---the embeddings, although generated by a neural model, are not ``tunable'' without investing in re-training the model, which is outside the scope of \tname{}. We searched approximately 200 configurations and found, on average, HDBSCAN was able to cluster \percentAverageClustered{}\%  (\numAverageClustered{}) of the \numFilesBroken{} broken \dfile{}s identified by \tname{}. In the clustering that we used, consisting of \numClusters{} clusters containing \numFilesInClusters{} files, we were able to confirm that 36.5\% of the clusters consisted of \dfile{}s that all had the same root cause for their failures.}

\begin{standout}
    \textbf{Summary of RQ2:} \tname{}'s approach to clustering \dfile{}s can, on average, cluster 34\% of our dataset, and, for over a third of the clusters generated, we can confirm that a \emph{single} issue covers all failing \dfile{}s within a cluster. 
\end{standout}

The answer to RQ2 bodes well for using clusters to bootstrap finding \added[id=J]{automated repairs. However, we note that the clustered files only make up a portion of broken files: therefore, to assess generalizability, RQ3 examines \tname{}'s ability to use repairs learned from our clustered files and apply them to non-clustered files.}

\subsection{Answering RQ3: \rqthree}

This question evaluates \tname{}'s effectiveness on our dataset of broken \dfile{}s (\Cref{sec:dataset}). 

RQ3.1 evaluates how much of the set of broken \dfile{}s can be addressed with the \added[id=J]{repairs} that \tname{} generates. Figure~\ref{fig:proportion} shows the effects \added[id=J]{of the repairs (and suggestions)} that we found across the \numClusters{} clusters produced by \tname. Each vertical bar denotes one cluster. These bars are divided into three segments. The size of the segment at the bottom of the bar (in yellow) represents the percentage of failures in a given cluster for which \tname{} provided \added[id=J]{an automated repair};
the size of the segment in the middle  of the bar (in blue) represents the percentage of failures for which \tname{} provided suggestions \added[id=J]{(which are generated in cases where no repairs apply)};
and the size of the segment at the top of the bar (in gray) represents the percentage of failures for which \tname\ could \emph{not} find a solution.

\begin{standout}
   \added[id=J]{\textbf{Summary of RQ3.1:} \replaced[id=L]{The}{Using the} \numrepairs\ repairs created with \tname\ offered solutions to \numPercRepair\% of the \numFilesInClusters\ broken and clustered \dfile{}s. In cases where no repairs were applicable, \tname{}'s \numrecommendations\ suggestions applied to an additional \numPercRecommend\% of the broken and clustered \dfile{}s.}
\end{standout}


RQ3.2 evaluates the ability of the repairs to generalize to a large number of cases. 
\Cref{table:repair-coverage} shows the relative amount of broken \dfile{}s that each one of our \numrepairs{} repair patterns covered. 

\input{figs/repair-coverage-table}

Column ``Id'' refers to the id of the repair (most of which listed in \Cref{table:fixes-main-2}), 
and column ``\#Clusters'' shows the number of clusters where the corresponding repair could fix at least one of the broken \dfile{}s in it.
Error patterns are extracted from a given cluster, which we refer to as ``parent''. 
Column ``(Coverage)~Parent'' then shows the fraction of broken \dfile{}s within the parent cluster that were corrected using the respective repair.
Column ``(Coverage)~Avg.'' shows the average fraction of repaired files across the different clusters affected by a repair pattern. 

\begin{standout}
    \textbf{Summary of RQ3.2:}~The \numrepairs\ repair patterns produced with \tname\ generalized well within the parent cluster (avg. 84.01\%) and across affected clusters (avg. 68.22\%).
\end{standout}

Recall that a total of \notInClusters{} of the \totalBrokenDockerfiles{} broken \dfile{}s (\notInClustersPerc{}\%) were \emph{not} clustered. \added[id=J]{For non-clustered files, \tname{} produced repairs to \numRepairNotInClustersPerc{}\% of them. Overall, \tname\ produced an actionable solution to the developer in \numCoverageNotInClustersPerc{}\% of the files that were not associated with any cluster (\numRepairNotInClustersPerc{}\% from repairs and an additional \numRecommendationsNotInClustersPerc{}\% from suggestions)}. Note that \tname\ used the patterns produced by analyzing \added[id=J]{clustered files}. That was possible because the clustering step is conservative and clusters \added[id=J]{were} based on embeddings of largely syntactic information (logs). For example, we observed that a file failing on the statement \CodeIn{apk add A \&\& apk add B \&\& ... \&\& apk add bzr} was not clustered with other files failing on \CodeIn{apk add bzr}---but, upon further examination, we found that this file failed to cluster due to its use of a conjunction of successive \CodeIn{apk add} commands instead of the (more common) use of the multi-argument \CodeIn{apk add A B ... bzr} variant. In practice, although conservative, the generated clusters were suitable for creating useful and generalizable repairs.

Finally, even when no repairs or suggestions apply, \tname{} can still provide a list of URLs pointing to resources that may provide a developer with a fix for their broken file. 



\begin{standout}
    \textbf{Summary of RQ3.3:} 
    \added[id=J]{Even in non-clustered broken \dfile{}s, \tname{} was able to produce automated repairs in \numRepairNotInClustersPerc{}\% of the files. Furthermore, when no repairs applied, \tname{} was able to provide suggestions in \numRecommendationsNotInClustersPerc{}\% of the files.}
\end{standout}

\subsection{Answering RQ4: \rqfour}
\label{sec:usefulness}

This section reports on two experiments we conducted to assess the practical usefulness of \tname. The first experiment (\Cref{sec:repair-confirmation}) measures the fraction of initially-broken but later-fixed \dfile{}s that could have been repaired with \tname. The second experiment (\Cref{sec:pull-requests}) measures the acceptance ratio of Pull Requests (PRs) for \dfile{}s found to be still broken in their repositories.

\subsubsection{Repair Confirmation}
\label{sec:repair-confirmation}


This experiment evaluates \tname\ on real patches created by \github{} developers. The metric we used was the fraction of the patches created by developers that matched the repairs or suggestions of \tname. To run this experiment, we searched for fixed \dfile{}s on \github. We used the same procedure as reported in Section~\ref{sec:data-collection}, but we re-cloned the repositories on Aug. 14, 2020 (8/14/20).
Because we know that the \dfile{} build on the first version of the project failed, we only needed to perform \dfile{} builds for the 8/14/20 versions of projects. To avoid unnecessary builds, we looked for \dfile{}s that were changed in the repository, and found that \numFixedFilesFromOriginallyBroken~(=8.87\%) of the original \numFilesInClusters\ broken \dfile{}s were changed in their repositories from the day they were retrieved up to 8/14/20.
We ran the command \CodeIn{docker build} in-context on those \numFixedFilesFromOriginallyBroken\ files, and discarded the cases where the build was still unsuccessful. In the end, we obtained a set of \numFoundFixes~$\langle{}x,y\rangle$ pairs to analyze, with $x$ denoting a broken \dfile{} from our dataset and $y$ denoting its corresponding patch. The method we used to measure effectiveness of \tname\ was to run \tname{} on $x$ and compare the generated repair or suggestion, if found, with $y$.


Of the \numFoundFixes\ cases of initially-broken then-fixed \dfile{}s, \tname\ produced an identical repair in \numMatches\ of the cases. In 77 cases, \tname\ provided suggestions that matched the patch used by the developer. Although we found that the ratio of suggestions to fixes was higher compared to results of RQ3.1, \tname\ covered most of the cases we analyzed (a total of \percCoverageRepairConfirmation\% of the cases). Overall, we believe that this result is encouraging because it provides a strong (and relatively unbiased) indication that the repairs that \tname\ produces are (i) correct (they matched the fixes of developers) and (ii) useful (almost all cases were covered).


\subsubsection{Pull Requests (PRs)}
\label{sec:pull-requests}

This experiment evaluates \tname\ on Pull Requests (PRs) issued to \github{} projects with still-broken \dfile{}s. The goal is to assess the feedback from developers to these PRs, which is a proxy for their interest in \tname{}'s results. 
For each of the \numrepairs{} repair patterns, we randomly sampled 5 \dfile{}s (from our dataset) that remained broken until the date we ran this experiment. Then, we manually prepared a PR that explained the problem (including a link to a similar case) and proposed a repair, as created by \tname. \deleted[id=J]{Table~\ref{tab:PRs} shows a summary of the \emph{accepted} PRs, including the link to the PR (Column ``URL''), and the ``Repair ID''.} To avoid violating double-blind rules, we created and used a \github\ account under the fictitious name ``Joseph Pett'' to submit the PRs. Our artifact (\artifactUrl{}) includes an up-to-date tracker of the submitted, accepted, and rejected PRs.

Of the \totalNumberOfPRs{} PRs that we submitted, \totalNumberOfPRsAccepted\ were accepted by developers (=\totalPectenOfPRsAccepted\%);
4 PRs were rejected;
and \NonReviewedPRs\ PRs have not yet been reviewed by developers.
The number of submitted PRs was lower than 65 (=\numrepairs*5) because we could not find five \dfile{}s still broken for some of the patterns. 

Three of the four rejected PRs were related to the same organization and the same problem, characterized by pattern \#7 (\Cref{table:fixes-main-2}). The developer pointed out that using a new version of the Docker Ruby image solved the encoding problem, and he preferred to update the Ruby version. With that feedback, we revised repair \#7 to include a second solution, which is to update the Ruby version to $2.5.8$. We have confirmed that this repair also works for the \dfile{}s repaired by the original solution. 
The new version of the Ruby image was committed on June 2020~\cite{ruby-encoding-image-update}, while this issue has been reported since June 2015~\cite{ascii-fix}. 
This GitHub issue was the URL recommended by \tname{} to assist the human to produce a repair.




\begin{standout}
    \textbf{Summary of RQ4}: 
    These  results  provide  initial,  yet strong,  evidence  that \tname\ is a useful aid to help developers fix broken \dfile{}s.
\end{standout}

\section{Threats to Validity}
\label{sec:threats}

Although most of our analysis is based on samples of (broken) \dfile{}s from GitHub repositories with ten or more stars, it is possible that this data is not representative of \dfile{} use in general.
Nonetheless, we found that real developers accepted the patches that \tname{} generated, and thus we can be reasonably sure that the trends (and repairs) we have identified are applicable in practice.
Additionally, our repairs and suggestions require a human in the loop, which is a source of bias.
To side-step this source of bias, we ran a ``time-travel'' analysis in which we were able to confirm retroactively that \tname{}'s repairs and suggestions were either identical (for repairs) or similar (for suggestions) to patches that developers actually applied.
This study is an important counterpoint to our pull-request study, because even pull-request acceptances could be biased, in the sense that it is hard for a developer to ``reject'' an offered patch.

We also made efforts to bolster our results by using robust methodology where possible:
e.g., to understand clustering behavior, we ran a grid search over 200 different configurations of hyper-parameters;
we also benchmarked two recent static tools to give some context to \tname{} results.

\section{Related Work}
\label{sec:related}


\noindent\textbf{Empirical studies on Docker (and DevOps).}~
A growing number of studies have been carried out on \dfile{}s, as well as on the broader topic of DevOps \cite{DBLP:journals/infsof/RahmanMW19} (also known as \emph{infrastructure as code}).
For Docker, Cito~\etal~\cite{CitoEmpiricalDockerEcoMSR2017} examined \dfile{} quality and, similar to us, found a high rate of breakage in \dfile{} builds; they cite a 34\% breakage rate from a smaller sample of 560 projects. We found a comparable breakage rate, but have also developed methods aimed at making \emph{repairs} instead of just analyzing quality. More recently, Wu~\etal{}~\cite{empirical-study-build-failures-docker} conducted a comprehensive study of build failures in \dfile{}s. They analyzed a total of 3,828 GitHub project containing \dfile{}s, and a total of 857,086 Docker builds. Overall, they found a failure rate of 17.8\%. Despite the differences in failures rates, these studies corroborate our finding that build failures are prevalent. Lin~\etal{}~\cite{dataset-docker-images} analyzed patterns (\ie{}, good and bad practices) in \dfile{}s. Among various observations, they found that many \dfile{}s use obsolete OS images, which can pose security risks (because attackers could exploit documented vulnerabilities) and incorrectly use the latest tag. Xu and Marinov~\cite{XuMiningContainerICSE2018} investigated characteristics of Docker images from DockerHub. Among other findings, they listed opportunities to improve Software Engineering tasks based on how images are organized. For example, they report that image variants could be used to support combinatorial testing. Zerouali \etal{}~\cite{ZeroualiOutdatedContainersSANER2019} studied version-related vulnerabilities (yet another category of issues that may arise in \dfile{}s---similar to some of the build-breaking issues we observed, in which \emph{external changes in the environment} negatively effect a \dfile{}). Among various findings, they found that no release is devoid of vulnerabilities, so deployers of \docker{} containers cannot avoid vulnerabilities even if they deploy the most recent packages. 

\noindent\textbf{Analysis of \dfile{}s}~
Henkel \etal~\cite{henkel2020learning} created a static checker for \dfile{}s (similar to Hadolint \cite{github:Hadolint}), called \binn{}, which is capable of learning rules from existing \dfile{}s;
however, unlike \tname{}, neither \binn{} nor Hadolint attempts \emph{repairs}.
Xu \etal~\cite{XuDockerfileTFSmellCOMPSAC2019} examined ``Temporary File Smells'', which are an \emph{image-quality}-related issue, not a \emph{build-breaking} issue, such as the ones we examined. Zhang \etal~\cite{zhang2018insight} studied the effect of \dfile{} changes on build time and quality (and utilized the static tool Hadolint). Hassan \etal{}~\cite{rudsea} proposed RUDSEA a tool-supported technique that proposes updates in \dfile{}s. RUDSEA analyzes changes in software environment assumptions---obtained with static analysis---and their impacts. We consider RUDSEA and \tname{} to be complimentary approaches: RUDSEA focuses on changes within a project and \tname{} focuses on changes external to a project. Other empirical studies on DevOps, but not \docker, include an examination of \emph{smells} in software-configuration files~\cite{sharma2016does}, and a study of the coupling between infrastructure-as-code files and ``traditional'' source-code files~\cite{jiang2015co}.

\noindent\textbf{Automated Code Repairs.}~
\tname{} lies within the growing body of work in automated repair.
According to a recent survey~\cite{automatic-sw-repair-survey}, our approach can be classified as both \emph{Generate-and-Validate} and \emph{Fix Recommender}. We use pre-defined templates that are obtained (i) via the analysis of build logs extracted from our clusters, and (ii) from examples found in community websites.
As such, we side-step the challenge of a fully automatic repair process to produce acceptable fixes. In addition to automated repair of source code, there is a growing effort to automate repair of build-related (DevOps) artifacts. These DevOps artifacts are unique in that they are often tied to both a source repository and the broader external environment in which one wants to build, test, and/or run their code. In the broader context of repair for build-related artifacts, both Lou \etal{}~\cite{history-driven-build-failure-fixing} and Hassan and Wang~\cite{HireBuild} investigate repair in the domain of Gradle builds (a kind of DevOps artifact used in many Java projects) and Macho \etal{}~\cite{BuildMedic} explore the related problem of automated repair for Maven builds.

\noindent\textbf{Broken Updates in Package Managers.}~
Prior work investigated the impact of breaking changes in package managers.
Mancinelli \etal{}~\cite{4019575} formalized package dependencies within a repository, and encoded the installability problem as a SAT problem.\deleted[id=M]{
Their focus was on distribution editors looking to improve their package repositories.} 
Vouillon and Cosmo~\cite{6606587} proposed an algorithm to identify \emph{broken sets} of packages that cannot be upgraded together within a component repository. 
McCamant and Ernst~\cite{mccamant-ecoop2004} proposed an approach for checking incompatibility of upgraded software components. 
They compute operational abstractions based on input/output behaviour to test whether a new component can replace an old one. 
M\o{}ller and Torp~\cite{breaking-changes-node} proposed a model-based testing approach to identify type-regression problems that result in breaking changes in JavaScript libraries.
These works deal with improvements and repairs applied to a package repository or library, and thus have a different focus compared to \tname{}, which is on repairing \emph{broken} \dfile{}s.
%
More related to checking inconsistencies\replaced[id=M]{of}{on} client code, Tucker \etal{}~\cite{4222580} proposed the OPIUM package-management tool.
Given a set of installed packages and information about dependencies and conflicts, they used a variety of solvers to determine (i) if a new package can be installed; (ii) the optimal way to install it; and (iii) the minimal number of packages (possibly none) that must be removed from the system.
\tname{} does not rely on explicit information about dependencies (which might not be available or feasible to obtain).
Instead, it extracts information from build logs, and leverages community knowledge bases to find solutions.
This approach enables \tname{} to address problems that go beyond broken packages and conflicts.





\section{Conclusions}
\label{sec:conclusion}

In an analysis of many open source repositories that use Docker, we found a surprising number of \emph{broken} \dfile{}s, most of which existing static analyzers cannot detect. For the cases they can detect, they do not propose \added[id=J]{repairs}. To address this problem, we propose \tname{}, a \added[id=J]{human-in-the-loop} approach for clustering, analyzing, and fixing broken \dfile{}s\deleted[id=J]{(through repairs and suggestions)}. We conducted a comprehensive evaluation of \tname, which showed that it was a helpful aid to fix broken \dfile{}s on \github{}. \added[id=J]{Using \tname\, we were able to submit \totalNumberOfPRs{} pull requests, of which \totalNumberOfPRsAccepted{} were accepted.} \added[id=J]{Furthermore, the tools and data we produced as result of this study are publicly available online via our artifact: 
\begin{center}\artifactUrl{}\end{center}.}

\section*{Acknowledgments}
\thanks{Supported, in part, by a gift from Rajiv and Ritu Batra;
by Facebook under a Probability and Programming Research Award;
and by ONR under grants N00014-17-1-2889 and N00014-19-1-2318; by INES 2.0, FACEPE grants PRONEX APQ 0388-1.03/14 and APQ-0399-1.03/17, CAPES grant 88887.136410/2017-00, FACEPE grant APQ-0570-1.03/14 and CNPq (grants 465614/2014-0, 309032/2019-9, 406308/2016-0, 409335/2016-9).
Any opinions, findings, and conclusions or recommendations
expressed in this publication are those of the authors,
and do not necessarily reflect the views of the sponsoring
entities.}


\balance
\bibliographystyle{IEEEtran}
\bibliography{IEEEabrv,references,devops}
    
\end{document}

%% file: packages.tex
\usepackage{pgfplots}
\pgfplotsset{width=7cm,compat=1.8}
\usepackage{balance}
\usepackage[font=footnotesize]{caption}
\usepackage[T1]{fontenc}
\usepackage{colortbl}
\usepackage{multirow}
\usepackage{subcaption}
\usepackage{listings}
\usepackage{comment}
\usepackage{adjustbox}
\usepackage{booktabs}
\usepackage[inline]{enumitem}
\usepackage{ textcomp }
\usepackage{wrapfig}
\usepackage{tabularx}
\usepackage{csvsimple}
\usepackage{hyperref}
\usepackage{cleveref}
\usepackage[framemethod=tikz]{mdframed}

\usepackage{savesym}
\savesymbol{comment}
\usepackage[final,authormarkup=none]{changes} %

\definechangesauthor[name=Jordan, color=blue]{J}
\definechangesauthor[name=Marcelo, color=red]{M}
\definechangesauthor[name=Leopoldo, color=purple]{L}
\definechangesauthor[name=Tom, color=green]{T}
\definechangesauthor[name=Denini, color=yellow]{D}

\usepackage{color}
\definecolor{LightGray}{rgb}{0.75,0.75,0.75}
\definecolor{VeryLightGray}{rgb}{0.90,0.90,0.90}
\definecolor{ForestGreen}{rgb}{0.13,0.54,0.13}

\usepackage{cite}
\usepackage[noend]{algpseudocode}
\usepackage{algorithm}
\usepackage{xcolor}
\usepackage{xstring}
\usepackage{tikz}
\usetikzlibrary{calc,matrix,chains,positioning,decorations.pathreplacing,arrows}
\usepackage[skins]{tcolorbox}
\usepackage{pgf-pie}

\definecolor{commentgreen}{RGB}{2,112,10}
\definecolor{eminence}{RGB}{108,48,130}

\lstdefinestyle{Bash}{
language=bash,
morekeywords={FROM,RUN,ARG,ADD},
showspaces=false,
basicstyle=\ttfamily,
commentstyle=\color{commentgreen},
keywordstyle=\color{eminence},
basicstyle=\scriptsize, 
numbers=left,
numbersep=5pt, 
stepnumber=1,
showstringspaces=false,
xleftmargin=3.0ex,
tabsize=1,
breaklines=true,
breakatwhitespace=false,
linewidth=8.65cm
}

\usepackage{xspace}

%% file: macros.tex
\newcommand{\CodeIn}[1]{\begin{footnotesize}\texttt{#1}\end{footnotesize}}

\newcommand{\Ignore}[1]{}
\newcommand{\myeg}{e.g.}
\newcommand{\ie}{i.e.}
\newcommand{\etc}{etc.}

\newcommand{\numrepairs}{13}
\newcommand{\numrecommendations}{50}

\newcommand{\numFixedFilesFromOriginallyBroken}{161} 
\newcommand{\numFoundFixes}{102} 
\newcommand{\numMatches}{23}
\newcommand{\numMatchesPerc}{22.77}
\newcommand{\numClusters}{144}
\newcommand{\numPercRecommend}{69.63}
\newcommand{\numPercRepair}{20.34}

\newcommand{\percCoverageRepairConfirmation}{98.04} 

\newcommand{\totalNumberOfPRs}{45}
\newcommand{\totalNumberOfPRsAccepted}{19}
\newcommand{\totalPectenOfPRsAccepted}{42.2}

\newcommand{\NumCasesOfIdFiveInDataset}{17}
\newcommand{\NonReviewedPRs}{\pgfmathparse{int(\totalNumberOfPRs-\totalNumberOfPRsAccepted-4)}\pgfmathresult}

\newcommand{\notInClusters}{3,586}
\newcommand{\notInClustersPerc}{66.4}
\newcommand{\totalBrokenDockerfiles}{5,400}

\newcommand{\numRepairNotInClustersPerc}{18.18}
\newcommand{\numRecommendationsNotInClustersPerc}{46.63} 

\newcommand{\numCoverageNotInClustersPerc}{64.81} 

\newcommand{\coverageOfAllFiles}{18.9}

\newcommand{\coverageOfAllFilesIncludingSuggestions}{73.25}

\newcommand{\etal}{\textit{et~al.}\xspace}

\newcommand{\tcentered}[1]{\begin{tabular}{@{}c@{}} #1 \end{tabular}}

\newcommand{\Mar}[1]{[\textbf{Marcelo}:{\color{magenta} #1}]}
\newcommand{\Leo}[1]{[\textbf{Leopoldo}:~{\color{blue} #1}]}

\definecolor{amethyst}{rgb}{0.6, 0.4, 0.8}

\newcommand{\tname}{\textsc{Shipwright}}
\newcommand{\dfile}{Dockerfile}
\newcommand{\docker}{Docker}
\newcommand{\github}{GitHub}

\definecolor{dkgreen}{rgb}{0,0.6,0}
\definecolor{ltgray}{rgb}{0.5,0.5,0.5}

\makeatletter
\newif\ifcolname
\colnamefalse

\def\keywordcheck{%
\IfStrEq*{\the\lst@token}{select}{\global\colnametrue}{}%
\IfStrEq*{\the\lst@token}{where}{\global\colnametrue}{}%
\IfStrEq*{\the\lst@token}{from}{\global\colnamefalse}{}%
\color{blue}%
}
\def\setidcolor{%
\ifcolname\color{purple}\else\color{black}\fi%
}
\makeatother

\newcommand{\Space}[1]{}
\newcommand{\Contrib}[1]{\textbf{#1}}

\newcommand{\binn}{\texttt{binnacle}}

\newcommand{\numTotalFiles}{32,466}
\newcommand{\numFilesBuilt}{20,526}
\newcommand{\numFilesBroken}{5,405}
\newcommand{\breakageRate}{26.3\%}
\newcommand{\numFilesTimedout}{393}
\newcommand{\numFilesUndet}{3,514}

\newcommand{\numFilesInClusters}{1,814}

\newcommand{\binnacleMaybeCoversP}{20.6}

\newcommand{\hadolintMaybeCoversP}{33.8}

\newcommand{\percentAverageClustered}{34}
\newcommand{\numAverageClustered}{1,836}
\newcommand{\artifactUrl}{\url{https://github.com/STAR-RG/shipwright}}
\newcommand{\toolUrl}{\url{https://github.com/STAR-RG/shipwright}}

\newmdenv[
    nobreak,
    middlelinewidth=.8pt,
    frametitlefont=\bfseries,
    leftmargin=0,
    rightmargin=0,
    skipabove=0.5\topsep,
    skipbelow=0.2\topsep,
    topline=false,
    bottomline=false,
    singleextra={
        \draw[line width=.8pt,black,] ($(O|-P)$) -- ($(O|-P)+(0.25cm,0)$); 
        \draw[line width=.8pt,black,] ($(O)$) -- ($(O)+(.25cm,0)$);
        \draw[line width=.8pt,black,] ($(P)-(0.25cm,0)$) -- ($(P)$); 
        \draw[line width=.8pt,black,] ($(P|-O)-(.25cm,0)$) -- ($(P|-O)$);
    },%
]{standout}


%% file: figs/fix-repairs-table.tex
\begin{table*}[t]
\caption{Selected \deleted[id=J]{Fix} Repairs.}
\label{table:fixes-main-2}
\resizebox{\textwidth}{!}
{
    \begin{tabular}{cccc}
    \toprule
    Id & Pattern & Repair & Src. \\
    \midrule
    \begin{filecontents*}{data/fixes-main-table-repairs.csv}
id, cluster, pattern, fix, source
1, 63-97-98-99-137, {``\textit{ERROR: Error installing bundler} $\in$ \underline{log} $\wedge$\\``\textit{bundler requires Ruby version} $\geq$ ([0-9]+.[0-9]+.[0-9]+)'' $\in$ \underline{log}}, {\underline{replace} \CodeIn{FROM ruby:}(.*) \underline{with} \CodeIn{FROM ruby:\$0}}, {\cite{bundler-fix}}
2, 57-65-66, {``\textit{Rpmdb checksum is invalid: dCDPT''} $\in$ \underline{log}}, {\underline{add} \CodeIn{RUN yum install -y yum-plugin-ovl} \underline{after} \CodeIn{FROM}(.*)}, {\cite{Rpmdb-fix}}
3, 55-63-69-85-94-107-113-114-115- \\ 116-119-120-129-130-131-132, {``\textit{E: Some index files failed to download.} \\ \textit{They have been ignored, or old ones used instead}'' $\in$ \underline{log}}, {\underline{replace} base image \underline{with} latest release from \url{hub.docker.com}}, {\cite{EOL-fix}}
4, 63-94-107, {``\textit{E: Package `libpng12-dev' has no installation candidate}'' $\in$ \underline{log}}, {\underline{replace} \CodeIn{libpng-12dev} \underline{with} \CodeIn{libpng-dev}}, {\cite{libpng-fix}}
5, 120-121, {``\CodeIn{FROM ubuntu}($\epsilon$|:latest|:20.04)'' $\wedge$\\``\textit{Unable to locate package} (.*)''$\in$ \underline{log} $\in$ \underline{\dfile}}, {\underline{replace} \CodeIn{\$0} \underline{with} \CodeIn{:18.04}, ...}, {\cite{pip-fix}}
6, 32-33, {``\textit{but your Gemfile specified} ([0-9\textbackslash{}|\textbackslash{}\textbackslash{}.]+)'' $\in$ \underline{log} \\$\wedge$ ``\CodeIn{FROM ruby}(.*)'' $\in$ \underline{\dfile}}, {\underline{replace} \CodeIn{FROM ruby:}(.*) \underline{with} \CodeIn{FROM ruby:\$0}}, {\cite{gemfile-fix}}
7, 125, {``\textit{invalid byte sequence in US-ASCII}'' $\in$ \underline{log} \\$\wedge$ ``\CodeIn{FROM ruby}(.*)'' $\in$ \underline{\dfile}}, {\underline{add} \CodeIn{ENV LANG C.UTF-8} \underline{after} \CodeIn{FROM}(.*)},{\cite{ascii-fix}}
8, 34-45-46-47-48-49-60-88-89, {``\textit{sh:} (.*)\textit{: not found}'' $\in$ \underline{log}}, {\textbf{if} ``\CodeIn{FROM alpine}(.*)'' $\in$ \underline{\dfile}\\\textbf{then} \underline{add} \CodeIn{RUN apk add --no-cache \$0}.\\\textbf{else} \underline{add} \CodeIn{RUN apt-get -y update }\\\CodeIn{\&\& apt-get -y install \$0}}, {\cite{notfound-fix}}
9, 70, {``\textit{ERROR: unsatisfiable constraints: bzr (missing):}'' $\in$ \underline{log} \\$\wedge$ ``\CodeIn{FROM alpine}(.*)'' $\in$ \underline{\dfile}}, {\underline{remove} \CodeIn{bzr} \underline{in} ``\CodeIn{apk add}'' command }, {\cite{bzr-fix}}
10,34-36,{``\textit{conda: not found}'' $\in$ \underline{log} \\$\wedge$ ``\CodeIn{RUN curl} (\textit{https://repo.continuum.}(.*))''  $\in$ \underline{\dfile}}, {\underline{add} \CodeIn{-L} \underline{before} \CodeIn{\$0}},{\cite{curl-fix}}
\end{filecontents*}
\csvreader[
    head to column names,
    late after line=\\ \vspace{0.5ex},
    late after last line=\\\bottomrule
    ]{data/fixes-main-table-repairs.csv}{}%
    {
      \tcentered{\id} & 
      \tcentered{\pattern} & 
      \tcentered{\fix} &
      \tcentered{\source} 
    }%
    \end{tabular}
}
\vspace{-1ex}
\end{table*}

%% file: figs/fix-recommendations-table.tex
\begin{table*}[t]
\caption{Selected \deleted[id=L]{Fix} Suggestions.}
\label{table:fixes-main-1}
\resizebox{\textwidth}{!}
{
    \begin{tabular}{ccc}
        \toprule
        Id & Pattern & Suggestion \\
        \midrule
        \begin{filecontents*}{data/fixes-main-table-suggestions.csv}
{id}, {pattern}, {fix}, {source}
1, {``\textit{mix}'' $\in$ \underline{log} $\wedge$ ``\textit{Code.LoadError}'' $\in$ \underline{log}}, {Problem running mix on the Elixir project...}, {-}
2, {``\textit{tsc}'' $\in$ \underline{log} $\wedge$ ``\textit{: error TS}'' $\in$ \underline{log}}, {Error during TypeScript compilation with tsc. Please check your .ts files.}, {-}
3, {``\textit{wget: server returned error}'' $\in$ \underline{log} $\vee$ ``\textit{wget: unable to resolve host address}'' $\in$ \underline{log}}, {\CodeIn{wget} error, you have a broken URL. Please check the log.}, {-}
4, {``\textit{npm ERR!}'' $\in$ \underline{log}}, {NPM Error, check your files, in particular, ``npm install'' commands.}, {-}
5, {``\textit{curl:}([0-9]+).*'' $\in$ \underline{log}}, {\CodeIn{curl} error, you have a broken URL. Please check the log.}, {-}
\end{filecontents*}
\csvreader[
        head to column names,
        late after line=\\ \vspace{0.5ex},
        late after last line=\\\bottomrule
        ]{data/fixes-main-table-suggestions.csv}{}%
        {
          \tcentered{\id} &
          \tcentered{\pattern} & 
          \tcentered{\fix} 
        }%
    \end{tabular}
}
\vspace{-3ex}
\end{table*}

%% file: figs/repair-coverage-table.tex
\begin{table}[h!]
\centering
\caption{\footnotesize\label{table:repair-coverage}Repair Coverage.}
    \begin{tabular}{ccrr}
    \toprule
    \multirow{2}{*}{Id} & \multirow{2}{*}{\# Clusters} & \multicolumn{2}{c}{Coverage (\%)}\\
    \cline{3-4}
    &  & Parent & Average \\
    \midrule
    \begin{filecontents*}{data/repair-coverage.csv}
id, covparent,covavg, cluster, numclusters
1, 100,61 ,63-97-98-99, 4
2, 40,25.67 ,57-65-66, 3
3, 100, 54,55-63-69-94-107-113-114-115, 8
4, 60,49.34 ,63-94-107, 3
5, 88,88 ,120-121, 2
6, 100,90.50 ,32-33, 2
7, 82.14,82.14 ,125, 1
8, 100, 88.67,34-45-46-47-48-60, 6
9, 100, 100,70, 1
10, 100,95.5 ,34-39, 2
11, 62,50,79-118-123,3
12,80,42,53-103,2
13,80,60,36-38-74,3
\end{filecontents*}
\csvreader[
    head to column names,
    late after line=\\ \vspace{0ex},
    late after last line=\\\bottomrule
    ]{data/repair-coverage.csv}{}%
    {
      \tcentered{\id} & 
      \tcentered{\numclusters} & 
      \tcentered{\covparent} &
      \tcentered{\covavg}
    }%
    \end{tabular}
    \vspace{-3ex}
\end{table}